# Semi-supervised learning for generalizable intracranial hemorrhage detection and segmentation

Emily Lin[a], BA; Esther L Yuh[a, *], MD, PhD

[a] Department of Radiology & Biomedical Imaging
University of California, San Francisco
185 Berry St, San Francisco CA 94107

* corresponding author

Manuscript Type: Original Research

SUMMARY STATEMENT

A semi-supervised learning paradigm used for intracranial hemorrhage detection and segmentation on head CT significantly improved model generalization capability on an out-of-distribution dataset.

KEY POINTS

- The examination-level AUC on an out-of-distribution dataset was significantly higher for the semi-supervised model (0.939) compared to the baseline supervised model (0.907) ($p = 0.009$).
- The Dice Similarity Coefficient (DSC) on an out-of-distribution dataset was significantly higher for the semi-supervised model (0.829) compared to the baseline supervised model (0.809) ($p = 0.012$).

## ABSTRACT

### Purpose

To develop and evaluate a semi-supervised learning model for intracranial hemorrhage detection and segmentation on an out-of-distribution head CT evaluation set.

### Materials & Methods

This retrospective study used semi-supervised learning to bootstrap performance. An initial "teacher" deep learning model was trained on 457 pixel-labeled head CT scans collected from one U.S. institution from 2010-2017 and used to generate pseudo-labels on a separate unlabeled corpus of 25,000 examinations from the RSNA and ASNR. A second "student" model was trained on this combined pixel- and pseudo-labeled dataset. Hyperparameter tuning was performed on a validation set of 93 scans. Testing for both classification (n=481 examinations) and segmentation (n=23 examinations, or 529 images) was performed on CQ500, a dataset of 481 scans performed in India, to evaluate out-of-distribution generalizability. The semi-supervised model was compared with a baseline model trained on only labeled data using area under the receiver operating characteristic curve (AUC), Dice similarity coefficient (DSC), and average precision (AP) metrics.

### Results

The semi-supervised model achieved statistically significantly higher examination AUC on CQ500 compared with the baseline (0.939 [0.938, 0.940] vs. 0.907 [0.906, 0.908]) ($p = 0.009$). It also achieved a higher DSC (0.829 [0.825, 0.833] vs 0.809 [0.803, 0.812]) ($p = 0.012$) and Pixel AP (0.848 [0.843, 0.853]) vs. 0.828 [0.817, 0.828]) compared to the baseline.

### Conclusion

The addition of unlabeled data in a semi-supervised learning framework demonstrates stronger generalizability potential for intracranial hemorrhage detection and segmentation compared with a supervised baseline.

## 1. INTRODUCTION

CT is the brain imaging modality most commonly used to diagnose acute traumatic brain injury. However, CT images can be challenging to interpret, as tiny abnormalities occupying as few as ~100 pixels in a noisy, low-contrast volume of $>10^6$ pixels must be detected. As CT images are characterized by image artifacts and low signal-to-noise, they are read by experts after years of training. Because even tiny, missed bleeds may have devastating consequences, the bar for machine learning models to be accepted into practice is exceedingly high. In addition, algorithms must be able to generalize even after attaining strong performance, maintaining high accuracy on scans produced by different scanners at different institutions.

In recent years, computer vision advancements have led to algorithms that can detect (1-4) and segment intracranial hemorrhage (5). The detection capability of such models can help streamline the triage process in radiologic workflow, while the segmentation capability can provide information about lesion localization and size, which are critical factors in downstream clinical management decisions. Segmentations can also be used for lesion quantification, which can be used to answer important questions about brain injury outcomes. A recent study demonstrated PatchFCN (6), a fully convolutional neural network, trained on $\sim10^5$ images with pixel-level reference standard segmentations, that achieved expert-level examination classification accuracy when tested on within-institution data. However, it is unclear whether such models can maintain expert-level accuracies on data from diverse sources. As variations in CT scanner hardware and technical parameters across institutions produce images with differing noise characteristics and image artifacts, models trained on data from one institution typically suffer performance losses when applied to data from other institutions. Therefore, improving a model's generalization capability is critical for it to be widely deployed.

To address this issue, we propose the use of semi-supervised learning, a machine learning strategy that uses both labeled and unlabeled data at training time. Noisy Student (7), a semi-supervised learning paradigm, demonstrated state-of-the-art performance on ImageNet, a widely used computer vision benchmark. Although semi-supervised learning has demonstrated promise on a variety of medical imaging applications (8-9), it has not yet been applied for the purpose of improving the generalizability of brain hemorrhage detection models (4, 10) on out-of-distribution datasets. In this study, we applied the semi-supervised Noisy Student learning paradigm (7) to detect and segment intracranial hemorrhage on head CT images. We demonstrate that allowing a model to learn from a broader complement of both labeled and unlabeled data improves its generalizability and performance on out-of-institution datasets.

## 2. MATERIALS AND METHODS

All CT examinations were collected retrospectively in Digital Imaging and Communications in Medicine (DICOM) format. Based on U.S. regulation 45 CFR 46.116(d) and the U.S. Food & Drug Administration, this study satisfied conditions for ethically acceptable waiver of patient consent and was approved by the institutional review board at the authors' institution. The study protocol was approved by the UCSF Committee on Human Research and is HIPAA-compliant. All segmentations were performed using an in-house custom graphical user interface application written in Python created by our group.

### 2.1 Datasets

The "Atlantis" labeled dataset consists of 457 pixel-labeled clinical head CT scans performed from 2010 to 2017 on 4 64-detector-row CT scanners from a single vendor at a single institution. It contains the typical spectrum of pathoanatomic types of intracranial hemorrhage and image artifacts seen in clinical practice, with skull, scalp and face removed for anonymity. In the Atlantis training dataset, 26.9% (123/457) of examinations are positive. Although this positivity rate is higher than that in the real world, it satisfies the need to include a sufficiently large sample of positive examinations and positive pixels. By hemorrhage subtype, 18.2% (83/457) of examinations contained subarachnoid hemorrhage, 14.0% (64/457) contained subdural hematoma, 13.3% (61/457) contained contusion, 5.0% (23/457) contained epidural hematoma, and 1.5% (7/457) contained intraventricular hematoma. Two board-certified neuroradiologists with 15 and 10 years of experience annotated all areas of hemorrhage at the pixel level. In an earlier study, data from all 457 CT examinations contained in the Atlantis dataset was also used to train PatchFCN (6). However, the goal of this prior study was to evaluate the classification accuracy of the model benchmarked against board-certified radiologists, while the present study seeks to use semi-supervised learning to improve model generalizability on out-of-distribution datasets.

The unlabeled training dataset, "Kaggle-25K", was curated by the Radiological Society of North America (RSNA) and the American Society of Neuroradiology (ASNR) and consists of an external corpus of >25,000 head CT examinations from the Kaggle/RSNA Intracranial Hemorrhage Detection competition (11). Kaggle-25K contains image-level labels but was treated as an unlabeled dataset for the purpose of semi-supervised learning. Data annotators can be found in (11).

For model development, we collected a validation dataset for hyperparameter tuning consisting of 93 head CT scans performed at hospitals represented in neither the labeled nor unlabeled datasets. 10.8% (10/93) of scans were positive. All scans were read by E.Y., a board-certified neuroradiologist with 14 years of experience.

For testing, we used the CQ500 dataset (2), curated by Qure.ai and the Center for Advanced Research in Imaging, Neurosciences and Genomics in New Delhi, India and licensed under a Creative Commons Attribution-NonCommercial-ShareAlike 4.0 International License and End User License Agreement. This public dataset includes 491 examinations, with reads from three radiologists with 8, 12, and 20 years of experience. We used the majority vote of the three radiologists to produce reference standard examination-level labels. The CQ500 test set contained 41.8% (205/491) positive examinations. The full spectrum of acute intracranial hemorrhage was included in CQ500, with 27.3% (134/491) of examinations containing intraparenchymal hemorrhage, 12.2% (60/491) containing subarachnoid hemorrhage, 10.8% (53/491) containing subdural hematoma, 5.7% (28/491) containing intraventricular hemorrhage, and 2.6% (13/491) containing epidural hematoma. 10 examinations that were chest or abdominal CT examinations, were missing slices, or consisted of 0.6mm slices not designed for human interpretation were excluded. To evaluate pixel accuracy, we randomly selected and pixel-labeled a 23-examination subset (k = 529 images) that reflected the positive examination rate of the overall dataset. We annotated a subset because pixel-labeling the full dataset was prohibitively expensive and time-consuming. Results on CQ500 have previously been published by the dataset curators (2), but we used CQ500 as an out-of-distribution test set for the purpose of evaluating our model's generalization capabilities. The data can be accessed at: http://headctstudy.qure.ai/dataset.

### 2.2 Semi-Supervised Algorithm Development

**Overview.** We applied the Noisy Student (7) learning paradigm. The training workflow **(Figure 1A)** is as follows: 1) Train a teacher model on a small pixel-labeled dataset. 2) The trained teacher model generates pixel-level and image-level pseudo-labels, or predictions, on a large unlabeled dataset. 3) Rank pseudo-labeled images from high to low based on probability of hemorrhage. 4) Apply a threshold, setting all images above the threshold to positive and all remaining images to negative. 5) Train a student model on the combination of the small pixel-labeled dataset and the larger pseudo-labeled dataset.

This trained student model was then tested on the overall CQ500 dataset and the pixel-labeled CQ500 subset to evaluate both examination-level and pixel-level performances, respectively **(Figure 1B)**. The teacher model was trained from scratch, while the student model was initialized based on the weights of the teacher model as per the Noisy Student paradigm.

**Data augmentation.** One important element in the Noisy Student approach (7) is the use of data augmentation to help achieve optimal performance. We explored three augmentation strategies: adjustments of image contrast, head size, and head aspect ratio. To employ contrast adjustment, each image in the training set underwent power law transform, logarithmic correction, or no change (12-13), with a random ⅓ selection probability for each. To modify head size and aspect ratio, the lengths and widths of each head were independently adjusted. Final augmentation parameters were selected based on the values that yielded the best performance on the validation dataset.

**Ranker.** We incorporated a ranker to reduce false positive errors, a common error type (**Figure 2**). First, the teacher model generates hemorrhage prediction probabilities for each unlabeled image. The images from the unlabeled dataset are ordered from high to low based on the hemorrhage probability predictions. A threshold is applied based on the teacher's predictions, with images above the threshold positive and images below it negative. To select the ranker threshold for semi-supervised model training, a radiologist first inspected a small set of images from the unlabeled dataset at various percentile thresholds $C \in \{5, 10, 15, 20, 25, 30\}$. At the ideal threshold, images with probabilities above threshold contain true positives without false positives. After inspection, we selected a threshold of $C = 10$; therefore, only the 10% of images with the greatest probability of hemorrhage were considered positive. For positive images, we set pixels with confidences of $K > 0.7$ to positive, $K < 0.3$ to negative. Values in between are ignored and generate no loss during training. All other images were considered negative, with all pixel predictions set to 0.

**Implementation Details.** The semi-supervised model was developed on PatchFCN (6), a fully convolutional neural network with a Dilated ResNet-38 backbone architecture and was initialized randomly. The PatchFCN model employs two branches, a classification branch and a segmentation branch, which leverage image- and pixel-level losses respectively. At training time, the model takes a patch of an image as input. The classification branch is supervised by patch-level binary class labels, while the segmentation branch is supervised by pixel labels. The patch-level class labels are derived from the pixel labels by checking whether the patch contains any positive pixel. Examination-level labels are not used in segmentation model training, and are used for evaluation only. At inference time, PatchFCN takes a patch of an image as an input, and outputs a binary class label (hemorrhage vs no hemorrhage) and segmentation pixel probability scores. The student model trained for 600 epochs with a step number of 240, batch size of 16, and a crop size of 240. We used a mixing ratio of 0.6-to-0.4 of Atlantis-to-Kaggle-25K for each minibatch during training. More details about PatchFCN may be found here: https://doi.org/10.1073/pnas.1908021116. The model was developed in Pytorch based on the following publicly available code: https://github.com/fyu/drn. Experiments were performed on a single NVIDIA Tesla V100 GPU.

**2.3 Models**

The **baseline model** consists of the PatchFCN model trained on the Atlantis pixel-labeled dataset only. The **semi-supervised (SS) model** consists of the PatchFCN model trained on both the Atlantis and Kaggle-25K dataset using the Noisy Student paradigm. Kaggle-25K was treated as an unlabeled dataset, and pixel pseudo-labels for this data were generated through the Noisy Student process.

Although the primary point of comparison is between the baseline and SS models, we also included the oracle models as additional performance benchmarks. Oracle models are those that are trained with additional labels *beyond* what would normally be available in a semi-supervised learning paradigm. The **image slice oracle model**, like the SS model, was trained on Atlantis and Kaggle-25K. However, it employs additional reference standard image labels that were curated and publicly available for Kaggle-25K, as this dataset did not come with pixel-level labels. At training time, pixel loss is activated on Atlantis only. On Kaggle-25K, the whole image is used as a patch, and the reference standard image label is equivalent to the patch label. The **pixel oracle model** goes one step further. On Kaggle-25K, it is trained with both the reference standard image labels and the SS model's pixel pseudo-labels. Pixel loss is activated on both Atlantis and Kaggle-25K, and reference labels are used to suppress false positive predictions. If the ground truth image label is negative, all pixel pseudo-labels for that image are set to 0. If the ground truth image label is positive, the teacher model's pseudo-label predictions are used. However, in a real-world clinical setting, the vast majority of data are unlabeled and have neither image nor pixel labels. Therefore, while the pixel oracle model represents the upper bound of performance and provides insight to assess the best achievable performance given the available data and labels, the primary comparison is between the baseline and semi-supervised models.

The **Qure.ai model** was developed by the curators of CQ500. We used this as an additional performance benchmark.

**2.4 Statistical Analysis**

The examination AUC reflects exam-level classification accuracy. To compare examination-level performance between the baseline and SS models, we collected 1000 sets of 481 CT examinations each, by conducting random sampling with replacement on the overall CQ500 dataset. We obtained $\Delta$AUC, the estimated differences in Examination AUC between the baseline and SS models, on all 1000 bootstrapped sets of 481 CT exams, and confirmed that the $\Delta$AUC are normally distributed at a confidence level of $p < 0.05$ using a standard Kolmogorov-Smirnov test for normality. We then calculated test statistic $Z = \Delta AUC / \sigma$, where $\sigma$ = the empirical standard deviation of $\Delta$AUC on the 1000 bootstrapped sets of 481 CT exams, and calculated a two-sided p-value as twice the probability that a standard normal random variable $< |Z|$. We also reported the 95% confidence interval of the $\Delta$AUC between models as the 2.5$^{th}$ and 97.5$^{th}$ percentile values of $\Delta$AUC.

To compare pixel-level performance between the baseline and SS models, we collected 1000 sets of 529 images, by conducting random sampling with replacement on the images comprising the pixel-labeled CQ500 test set. We obtained $\Delta$DSC, the estimated differences in DSC between the baseline and SS models on all 1000 bootstrapped sets of 529 images, and confirmed that the $\Delta$DSC are normally distributed at a confidence level of $p < 0.05$ using a standard Kolmogorov-Smirnov test for normality. We then calculated test statistic $Z = \Delta DSC / \sigma$, where $\sigma$ = the empirical standard deviation of the DSC differences on the 1000 bootstrapped sets of 529 images, and calculated a two-sided p-value as twice the probability that a standard normal random variable $< |Z|$. We also reported the 95% confidence interval of $\Delta$DSC between models as the 2.5$^{th}$ and 97.5$^{th}$ percentile values of $\Delta$DSC. We obtained $\Delta$AP, the difference in Pixel average precision (AP) between baseline and SS models in similar fashion, and reported the 95% confidence interval of $\Delta$AP between models as the 2.5$^{th}$ and 97.5$^{th}$ percentile values of $\Delta$AP.

### 2.5 Intra-Rater Reliability

The DSC and Pixel AP (14) segmentation metrics provide a measure of the localization capability of the system. To evaluate the segmentation results more rigorously and verify the superiority of the SS model compared with the baseline, we performed test-retest segmentations using a single labeler. The same board-certified neuroradiologist segmented the same 23-examination test set (529 images) a total of three times. Each segmentation was performed without review of prior segmentations, with a washout period of at least 7 days between examinations and random reordering of examinations.

To quantitatively determine whether intra-rater variability affects our results, we calculated the Pixel AP and DSC for the baseline model on 50 bootstrapped samples on each of the 3 neuroradiologist segmentations. This process was repeated for the SS model.

## 3. RESULTS

### 3.1 Patient Demographics

The SS model was trained on both the Atlantis and Kaggle-25K datasets. For the former, the mean age of the patients was 53.4 +/- 17.1 years and 36.3% (166/457) of the patients were female. Patient demographics for Kaggle-25K were not released. Main results were presented on CQ500, which was collected in two batches, B1 (43.9% (94/214) female; mean age, 43.4 years) and B2 (30.3% (84/277) female; mean age, 51.7 years).

### 3.2 Model Performance Comparison on CQ500

The SS model achieved a significant performance gain over the baseline model on the CQ500 test set across AUC, DSC and pixel AP metrics, demonstrating both superior localization and exam-level decision capabilities over the baseline **(Table 1)**. On the detection metric, the SS model demonstrates an examination AUC of 0.939 (95% CI: [0.938, 0.940]), performing significantly better than the baseline examination model, which had an AUC of 0.907 (95% CI: [0.906, 0.908]) ($p < 0.001$). The Qure.ai model (2), which yields state-of-the-art performance on CQ500, achieves an examination AUC of 0.941. In addition, using an operating point on the AUC curve at which the sensitivity is 0.920 on the CQ500 test set, the baseline and SS models showed a specificity of 0.630 and 0.804, respectively. The SS model also surpassed performance of the Image Slice Oracle model (0.894) ($p < 0.001$) and nearly matched that of the Pixel Oracle (0.943).

The DSC of the SS model was 0.829 (95% CI: [0.825, 0.833]), a significant improvement over the baseline DSC of 0.809 (95% CI: [0.803, 0.812]) ($p < 0.001$). The SS model demonstrated a Pixel AP of 0.848 (95% CI: [0.843, 0.853]), a significant improvement over the baseline Pixel AP (0.828 (95% CI: [0.817, 0.828])) as well the Image Slice Oracle AP (0.823 (95% CI: [0.843, 0.853]) and a near match to the Pixel Oracle AP (0.850).

Corresponding to the results on the metrics above, the SS model demonstrated increased ability to detect subtle bleeds and reject false positives. **Figure 3** shows baseline and SS model prediction visualizations on the validation set. The SS model was more robust against false positives, recognizing many of the baseline model's false positive predictions as true negatives **(Figure 3A, 3B)**. It was also able to accurately detect bleeds that the baseline model missed (**Figure 3C, 3D**).

**3.3 Data augmentation parameter optimization.** To identify the best data augmentation strategy, we investigated model performance using different contrast parameters (**Table 2**). Because Kaggle-25K pseudo-labels are generated at the pixel level, we optimized for Pixel AP. The parameters were fine-tuned on the baseline and evaluated on the validation set. The baseline Pixel AP was 0.651. We then implemented power law and logarithmic transforms individually, which yielded Pixel AP values of 0.653 and 0.649 respectively. Finally, we studied whether both strategies produced an additive effect by combining power law, log transform, and no augmentation with equal sampling ratios. Out of all contrast augmentation schemes, the combination strategy yielded the best pixel level performance (0.681) on the validation set. Therefore, we selected this combined data augmentation strategy as the one to use for the SS model.

**3.4 Ranker ablation study. Table 3** presents the results of our control study investigating the importance of a ranker for the SS model.

If the ranker is removed, the examination AUC drops from 0.948 to 0.918, as the model often generates pseudo-labels containing false positive predictions (**Figure 3**) that are suppressed by the ranker. However, without the ranker, the student would train on and reinforce its own false positive predictions.

**3.5 Test-retest reliability.** We found a statistically significant superiority of the SS model over the baseline model on each of the 3 neuroradiologist segmentations on both Pixel AP and DSC **(Table 4)**. Paired t-tests comparing the baseline and SS model on all 3 segmentations on Pixel AP and DSC showed p-values $< 0.001$.

## 4. DISCUSSION

One major barrier to the widespread clinical deployment of machine learning models is the need for improved generalization. Models must maintain high levels of performance accuracy on data from diverse sources encompassing heterogeneous populations and differing technical parameters. To address this, we employed a semi-supervised learning paradigm to detect and segment intracranial hemorrhage on head CT images, using a broad complement of both labeled and unlabeled data to improve model generalization capability. We demonstrated the potential for semi-supervised learning to significantly improve both granular and holistic localization capabilities. The examination AUC of the SS model (0.939) was significantly stronger than that of the baseline model (0.907). The DSC of the SS model was 0.829, which demonstrated a significant improvement over the baseline DSC of 0.809. The SS model demonstrated a Pixel AP of 0.848, a significant improvement over the baseline Pixel AP of 0.828. These results are qualitatively supported by the baseline and SS model output visualizations, in which the SS model demonstrated both increased sensitivity and specificity of predictions. The SS model (examination AUC, 0.939) was also able to achieve performance similar to that of Qure.ai (examination AUC, 0.941), the published state-of-the-art model on the CQ500 test set, despite using significantly more out-of-distribution data and 640-fold fewer labeled examinations.

The Image Slice Oracle and Pixel Oracle are provided as a means of comparison. The SS model likely outperforms the Image Slice Oracle because the pixel pseudo-labels on Kaggle-25K provide superior localization information. Although the Pixel Oracle performance minimally surpasses that of the SS model, it requires the cost of both the 285,000 extra image-level labels provided by experts. The SS model achieved very similar performance without the need for the time and monetary cost required to obtain these image-level labels.

We also demonstrated how important both data augmentation and the ranker are to maintaining strong performance. Data augmentation is helpful during training time because it exposes the model to a greater data distribution, thereby enhancing its generalization capability. CT scans may vary in contrast and in patient head sizes, so exposure to a wider data distribution would allow the model to learn from a broader and therefore challenging set of examples. The introduction of the ranker is also important. The majority of head CT scans are administered as cautionary screening tools and are therefore negative. The ranker operationalizes this prior clinical knowledge to minimize false positive errors, ultimately improving performance.

Because our goal was to evaluate generalizability, we selected the CQ500 test set which was curated on a different continent. This is a challenging setting for the model, as the data included scans performed on previously unfamiliar populations, with exposure to additional artifacts and technical characteristics that had not previously been encountered.

To our knowledge, semi-supervised learning has not been applied for generalizability of hemorrhage detection deep learning models. Salehinejad et al. (15) leveraged the RSNA Intracranial Hemorrhage CT dataset to explore the generalizability of their intracranial hemorrhage machine learning model. Remedios et al. (16) applied cyclic weight transfer to improve generalization for head CT hemorrhage segmentation. However, both approaches require all training data to be labeled and are unable to support the aggregation of labeled and unlabeled data as seen in semi-supervised learning. Additionally, neither model can be compared directly with ours due to differences regarding the choice of training and evaluation sets. Outside of these works, generalizability remains largely unexplored for this application. Wang et al. (17) has applied semi-supervised learning for intracranial hemorrhage segmentation, but they do not leverage this strategy for improved generalization. Many other works also exist to develop high-performing models for intracranial hemorrhage (2, 4, 6), but they do not tackle the issue of generalizability.

This study, however, still contains some limitations. In semi-supervised learning paradigms, the size of the unlabeled corpus is significantly larger than that of the labeled corpus. As a result, there is a greater computational requirement to process and train the large volumes of unlabeled data. As the amount of data increases, the required computational power and training time also increases. In addition, both labeled and unlabeled datasets typically grow with time in real-world clinical settings. The current workflow would require the algorithm to be completely re-trained from scratch with every increase in dataset size, which is a time-consuming process. To optimize this system for application, more research will be required to develop a semi-supervised learning scheme that requires minimal adaptation training on growing datasets.

In conclusion, we demonstrated that a semi-supervised learning paradigm could help achieve stronger generalization for intracranial hemorrhage classification and segmentation on head CT. We demonstrated the potential for algorithms to improve performance by aggregating both labeled and unlabeled data. While annotated data is valuable towards achieving strong performance, it also requires substantial time and energy to obtain. In contrast, unlabeled data can be found in abundance in clinical settings and is essentially “free” in its lack of requirement for any annotation time. Once an algorithm is seeded with a small high quality labeled dataset, a semi-supervised learning paradigm can raise the performance level simply by leveraging the vast unlabeled data in the real-world setting. The addition of unlabeled data through a semi-supervised approach can thus serve as a useful tool to improve generalizability and augment performance with minimal additional cost.

**5. References**

1. Titano, J.J., Badgeley, M., Schefflein, J., Pain, M., Su, A., Cai, M., Swinburne, N., Zech, J., Kim, J., Bederson, J. and Mocco, J., 2018. Automated deep-neural-network surveillance of cranial images for acute neurologic events. Nature medicine, 24(9), pp.1337-1341.
2. Chilamkurthy, S., Ghosh, R., Tanamala, S., Biviji, M., Campeau, N.G., Venugopal, V.K., Mahajan, V., Rao, P. and Warier, P., 2018. Deep learning algorithms for detection of critical findings in head CT scans: a retrospective study. The Lancet, 392(10162), pp.2388-2396.
3. Prevedello, L.M., Erdal, B.S., Ryu, J.L., Little, K.J., Demirer, M., Qian, S. and White, R.D., 2017. Automated critical test findings identification and online notification system using artificial intelligence in imaging. Radiology, 285(3), pp.923-931.
4. Lee, H., Yune, S., Mansouri, M., Kim, M., Tajmir, S.H., Guerrier, C.E., Ebert, S.A., Pomerantz, S.R., Romero, J.M., Kamalian, S. and Gonzalez, R.G., 2019. An explainable deep-learning algorithm for the detection of acute intracranial haemorrhage from small datasets. Nature Biomedical Engineering, 3(3), p.173.
5. Chang, P.D., Kuoy, E., Grinband, J., Weinberg, B.D., Thompson, M., Homo, R., Chen, J., Abcede, H., Shafie, M., Sugrue, L. and Filippi, C.G., 2018. Hybrid 3D/2D convolutional neural network for hemorrhage evaluation on head CT. American Journal of Neuroradiology, 39(9), pp.1609-1616.
6. Kuo, W., Häne, C., Mukherjee, P., Malik, J. and Yuh, E.L., 2019. Expert-level detection of acute intracranial hemorrhage on head computed tomography using deep learning. Proceedings of the National Academy of Sciences, 116(45), pp.22737-22745.
7. Xie, Q., Luong, M.T., Hovy, E. and Le, Q.V., 2020. Self-training with Noisy Student improves ImageNet classification. In Proceedings of the IEEE/CVF Conference on Computer Vision and Pattern Recognition (pp. 10687-10698).
8. Xie, Y., Zhang, J. and Xia, Y., 2019. Semi-supervised adversarial model for benign–malignant lung nodule classification on chest CT. Medical image analysis, 57, pp.237-248.
9. Masood, A., Al-Jumaily, A. and Anam, K., 2015, April. Self-supervised learning model for skin cancer diagnosis. In 2015 7th International IEEE/EMBS Conference on Neural Engineering (NER) (pp. 1012-1015). IEEE.
10. Arbabshirani, M.R., Fornwalt, B.K., Mongelluzzo, G.J., Suever, J.D., Geise, B.D., Patel, A.A. and Moore, G.J., 2018. Advanced machine learning in action: identification of intracranial hemorrhage on computed tomography scans of the head with clinical workflow integration. NPJ digital medicine, 1(1), pp.1-7.

11. Radiological Society of North America (RSNA). RSNA Intracranial Hemorrhage Detection dataset, retrieved September 2019 from https://www.kaggle.com/c/rsna-intracranial-hemorrhage-detection.
12. Yeo, I.K. and Johnson, R.A., 2000. A new family of power transformations to improve normality or symmetry. Biometrika, 87(4), pp.954-959.
13. Box, G.E. and Cox, D.R., 1964. An analysis of transformations. Journal of the Royal Statistical Society: Series B (Methodological), 26(2), pp.211-243.
14. Everingham, M., Van Gool, L., Williams, C.K., Winn, J. and Zisserman, A., 2010. The pascal visual object classes (voc) challenge. International journal of computer vision, 88(2), pp.303-338.
15. Salehinejad, H., Kitamura, J., Ditkofsky, N., Lin, A., Bharatha, A., Suthiphosuwan, S., Lin, H.M., Wilson, J.R., Mamdani, M. and Colak, E., 2021. A real-world demonstration of machine learning generalizability in the detection of intracranial hemorrhage on head computerized tomography. *Scientific reports, 11*(1), pp.1-11.
16. Remedios, S.W., Roy, S., Bermudez, C., Patel, M.B., Butman, J.A., Landman, B.A. and Pham, D.L., 2020. Distributed deep learning across multisite datasets for generalized CT hemorrhage segmentation. *Medical physics, 47*(1), pp.89-98.
17. Wang, J.L., Farooq, H., Zhuang, H. and Ibrahim, A.K., 2020. Segmentation of intracranial hemorrhage using semi-supervised multi-task attention-based U-net. *Applied Sciences, 10*(9), p.3297.

**Table 1.** Comparison of model performance on the CQ500 test set from India used to evaluate generalizability

| Model | CQ500 subset DSC | CQ500 subset Pixel AP | CQ500 Examination AUC |
|---|---|---|---|
| Baseline | 0.809 [0.803, 0.812] | 0.828 [0.817, 0.828] | 0.907 [0.906, 0.908] |
| **SS (Ours)** | **0.829 [0.825, 0.833]** | **0.848 [0.843, 0.853]** | **0.939 [0.938, 0.940]** |
| Image Slice Oracle | 0.787 [0.783, 0.787] | 0.823 [0.820, 0.823] | 0.894 [0.892, 0.894] |
| Pixel Oracle | 0.828 [0.827, 0.829] | 0.850 [0.849, 0.852] | 0.943 [0.942, 0.944] |
| Qure.ai | -- | -- | 0.941 |

Note — The CQ500 test set is a publicly available test set from India, which was used to evaluate model generalizability. The baseline model was trained on the Atlantis pixel-labeled dataset only. The semi-supervised (SS) model uses the semi-supervised learning technique shown in Figure 1 and was trained on both the Atlantis pixel-labeled and Kaggle-25K pseudo-labeled datasets (RSNA 2019). The Image Slice Oracle is a separate model trained on the same data as the SS model, but with image reference standard labels provided by Kaggle-25K curators (RSNA 2019). The Pixel Oracle represents the upper bound of performance, trained with the Kaggle-25K reference standard image labels as well as the SS model pseudo-labels. The Qure.ai model is listed for reference only, as it uses private training data and is the published state-of-the-art result on CQ500. AP = average precision, DSC = Dice similarity coefficient, AUC = area under the receiver operating characteristic curve.

**Table 2.** Study of data augmentation contrast parameters

| Contrast | Parameters | Pixel AP | Examination AUC |
| --- | --- | --- | --- |
| Baseline | -- | 0.651 [0.649, 0.651] | 0.965 [0.965, 0.967] |
| Power Law | $\gamma = U\,(0.85, 1.1)$ | 0.653 [0.651, 0.654] | 0.975 [0.974, 0.976] |
| Log | $g = U\,(0.7, 1.1)$ | 0.649 [0.647, 0.650] | 0.937 [0.935, 0.939] |
| **All** | See above | **0.681 [0.680, 0.682]** | 0.933 [0.929, 0.933] |

Note — We optimized the results for Pixel AP because contrast adjustment is low-level adjustment and Kaggle-25K pseudo-labels are generated through pixel-level segmentation. AP = average precision

**Table 3.** Control study demonstrating the importance of the ranker.

| Model | Validation Pixel AP | Validation Examination AUC |
| --- | --- | --- |
| No Ranker | 0.682 [0.680, 0.683] | 0.918 [0.914, 0.919] |
| **SS (Ours)** | **0.688 [0.686, 0.688]** | **0.948 [0.946, 0.949]** |

Note — Both the 'No Ranker' and 'SS (Ours)' model were performed without iterations in order to isolate the effects of the ranker. SS = semi-supervised

**Table 4.** Results of the test-retest study

| Test Set | Baseline Pixel AP | SS Pixel AP | Baseline DSC | SS DSC |
|---|---|---|---|---|
| Pixel-Labeled Set #1 | 0.828 [0.817, 0.82] | 0.848 [0.843, 0.853] | 0.809 [0.803, 0.812)] | 0.829 [0.825, 0.833] |
| Pixel-Labeled Set #2 | 0.854 [0.841, 0.855] | 0.874 [0.861, 0.874] | 0.824 [0.816, 0.825] | 0.842 [0.833, 0.842] |
| Pixel-Labeled Set #3 | 0.852 [0.844, 0.857] | 0.875 [0.868, 0.880] | 0.823 [0.817, 0.825] | 0.842 [0.836, 0.844] |

Note — On each of the 3 neuroradiologist segmentations, the semi-supervised model demonstrated a statistically significant increase in performance over the baseline. 95% confidence intervals are reported in brackets. AP = average precision, DSC = Dice similarity coefficient, SS = semi-supervised model

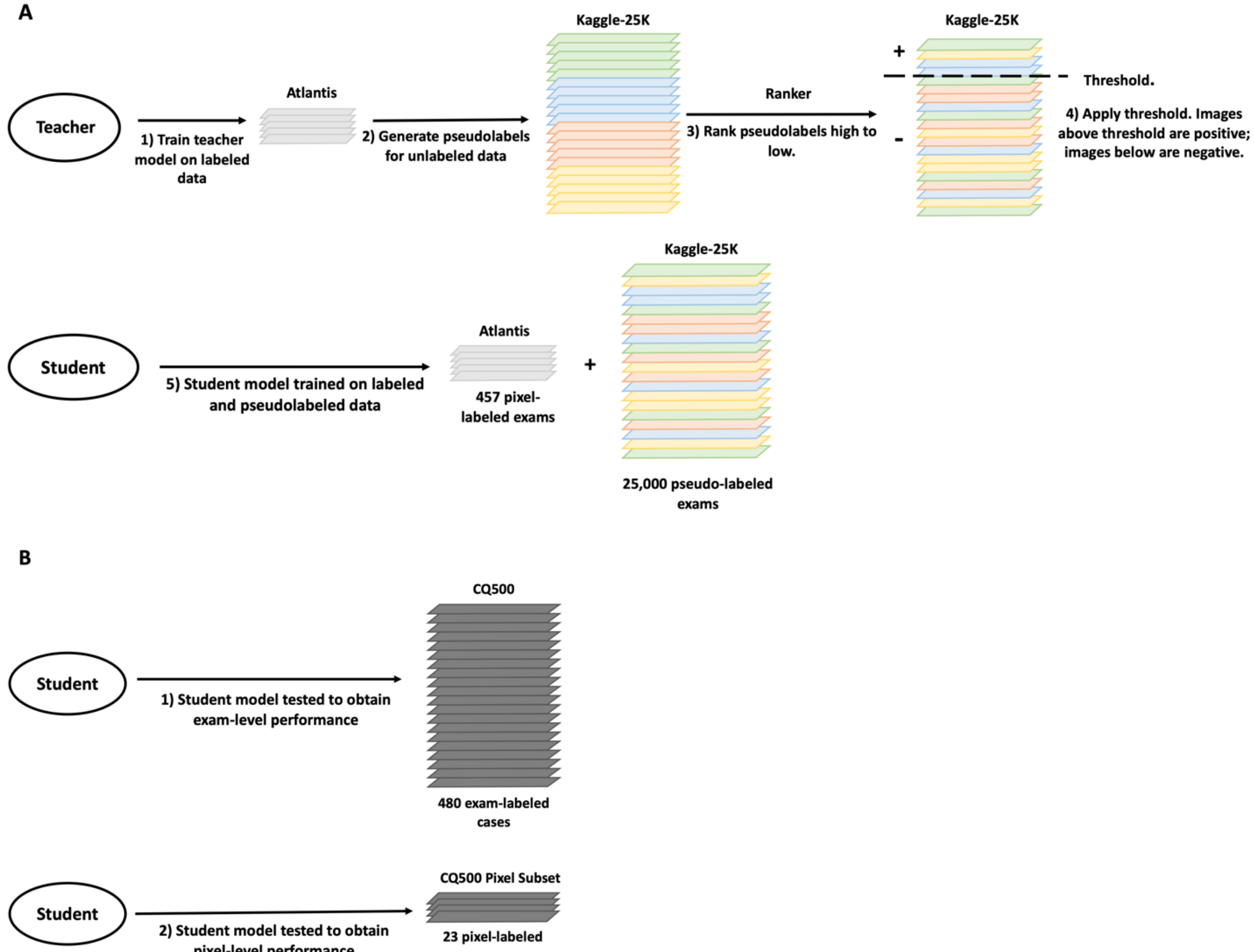

Figure 1. Schematic of the semi-supervised Noisy Student approach. Each color signifies data from a different institution. A) Workflow at training time, which is explained in further detail in Section 2.2. B) Workflow at test time as the student model is evaluated on both the CQ500 overall dataset and pixel-label subset to evaluate examination-level and pixel-level performances.

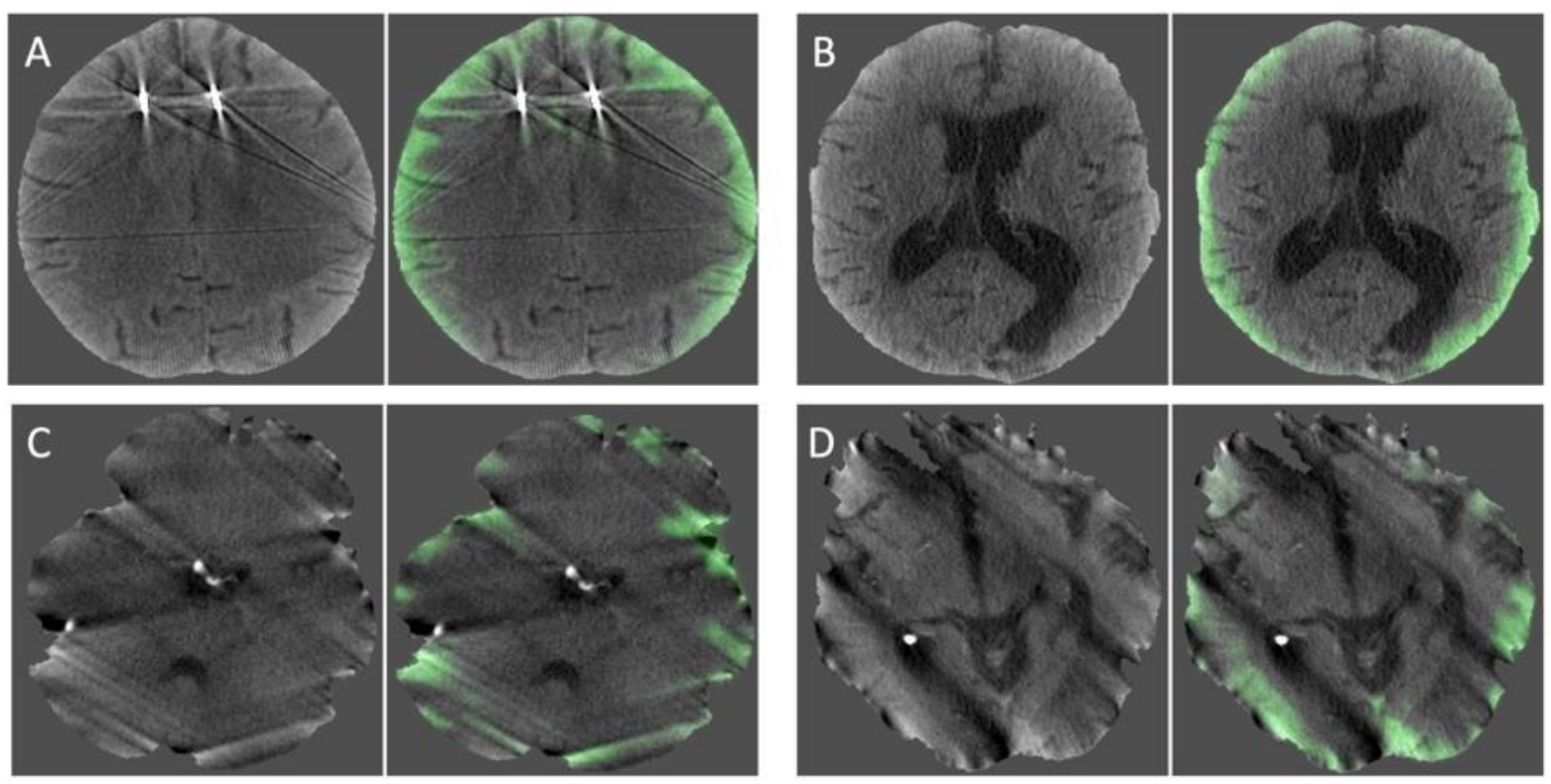

Figure 2. Visualization of false positive predictions, which are frequently made without the ranker. Green indicates the model predictions. These axial CT images obtained without contrast are from the validation set.

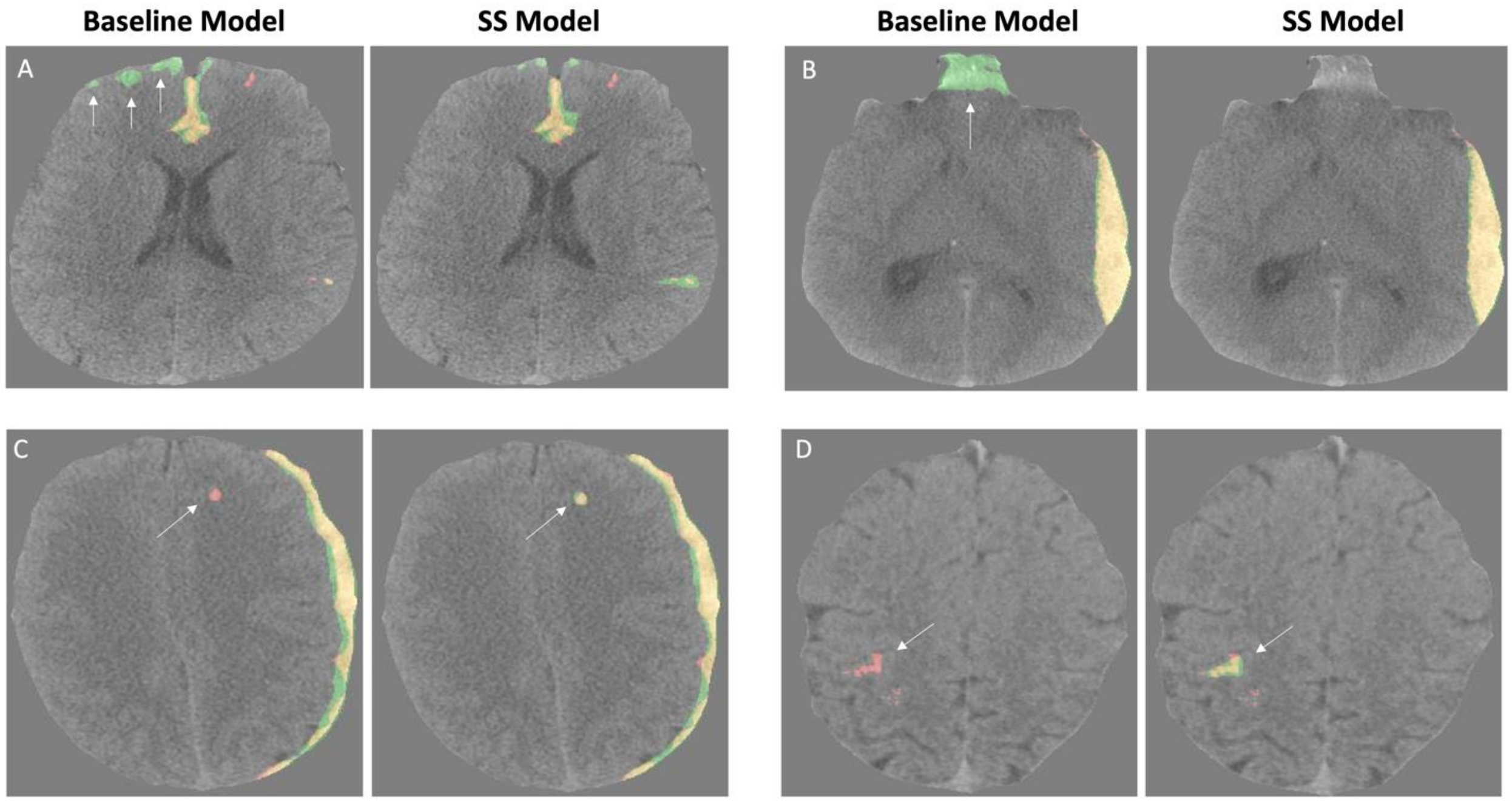

Figure 3. Visualization of model predictions on the validation set using the baseline and semi-supervised (SS) models. Red is the reference standard label, green is the model's positive prediction, and yellow is the overlap. These are axial CT images obtained without contrast.

| Intracranial Hemorrhage Subtype | Baseline Exam AUC | SS Exam AUC |
|---|---|---|
| Subarachnoid hemorrhage (SAH) | 0.914 | **0.936** |
| Subdural hemorrhage (SDH) | 0.829 | **0.919** |
| Intraparenchymal hemorrhage (IPH) | 0.872 | **0.894** |
| Epidural hemorrhage (EDH) | 0.922 | **0.976** |

**Supplemental Table 1.** Exam-level AUC for baseline and SS models by intracranial hemorrhage subtype. Because most head CT examinations contain multiple hemorrhage subtypes, we only considered cases with isolated hemorrhage subtypes to avoid confounding factors. AUC = area under the receiver operating curve.